# BEA-Base: A Benchmark for ASR of Spontaneous Hungarian

Péter Mihajlik[1,2], András Balog[1,3], Tekla Etelka Gráczi[1],
Anna Kohári[1], Balázs Tarján[2,3], Katalin Mády[1]

[1]Hungarian Research Centre for Linguistics, [2]Budapest University of Technology and Economics, [3]SpeechTex Inc.
[1]1068 Budapest, Benczúr u. 33., [2]1111 Budapest, Műegyetem rakpart 3., [3]1181 Budapest, Madách Imre utca 47. Hungary
mihajlik@tmit.bme.hu, {balog.andras, graczi.tekla.etelka, kohari.anna, mady}@nytud.hu, tarjanb@tmit.bme.hu

**Abstract**
Hungarian is spoken by 15 million people, still, easily accessible Automatic Speech Recognition (ASR) benchmark datasets – especially for spontaneous speech – have been practically unavailable. In this paper, we introduce BEA-Base, a subset of the BEA spoken Hungarian database comprising mostly spontaneous speech of 140 speakers. It is built specifically to assess ASR, primarily for conversational AI applications. After defining the speech recognition subsets and task, several baselines – including classic HMM-DNN hybrid and end-to-end approaches augmented by cross-language transfer learning – are developed using open-source toolkits. The best results obtained are based on multilingual self-supervised pretraining, achieving a 45% recognition error rate reduction as compared to the classical approach – without the application of an external language model or additional supervised data. The results show the feasibility of using BEA-Base for training and evaluation of Hungarian speech recognition systems.

## 1. Introduction

The development of Hungarian speech databases (Roach et al. 1996; Pollák et al., 2000; Siemund at al, 2000) and speech recognition research (Elenius & Takács, 1991; Gordos, 1991, Szarvas et al., 2000) has a long history. In practice, however, hardly any ASR result published for Hungarian is comparable to any other one in an exact way. Namely, most research groups used their in-house data without sharing them with other teams. Even if a public database was applied in a study, the selection and preprocessing of train/test sets was done usually individually, so the reconstruction of experimental setups and the reproduction of results may pose significant challenge. Thus, the lack of an easily accessible Hungarian language ASR benchmark dataset leads to a situation where speech recognition results for Hungarian are typically not directly comparable hindering the competition and the development of this important field.

A refreshing attempt was made recently by Mozilla launching the Common Voice (CV) project (Ardila et al. 2019) for Hungarian, but the data collection is still in an early stage, currently only 19 hours of verified training speech data is available. We look forward to the completion of the database, although only read sentences are going to be recorded. Conversational AI and similar real-life applications, however, are much better to evaluate on spontaneous speech – obviously no user will read his/her response to a virtual (or a real) agent.

Our aim was to create a benchmark dataset for Hungarian ASR training and evaluation, focusing on the speech-to-text component of conversational AI applications. Our main contributions are the following: We have selected the most concise part of the spontaneous language Hungarian BEA database (Gósy, 2013; Neuberger et al., 2014) – called BEA-Base in the rest of the paper – comparable in size to the WSJ dataset (Garafolo et al. 1993). We designated the training, development and evaluation sets and performed all preprocessing steps necessary for ASR experiments. We have developed and evaluated various baselines using open-source toolkits and public language resources. Finally, we made the benchmark dataset and the most relevant model checkpoints accessible through web[1] for the research community (after registration).

## 2. Related Work

### 2.1 Spontaneous Hungarian Speech Databases

One of the first Hungarian language conversational speech database was collected during the BUSZI (Budapest Sociolinguistic Interviews) project (Kontra 1997). Despite the considerable efforts made to record and transcribe the data, it is not accessible outside of the Hungarian Research Centre for Linguistics. Another renowned project was organized by the SHOAH foundation where about 2000 hours of Hungarian "Oral History" interviews were captured in the MALACH project (Oard et al, 2002; Mihajlik et al, 2007). Unfortunately, only a small fraction of the Hungarian speech is transcribed manually and it is not yet released to the public. The collection of BEA targeting mostly spontaneous speech (Gósy, 2013) began in 2007 and it still is in progress – the basic features of the database are summarized in Section 3.1. Next, we must mention the HuComTech Multimodal Database containing audio-visual recodings (about 60 hours) of 121 young adult speakers (Pápay et al., 2011). Unfortunately, none of these datasets could become a standard benchmark for spontaneous Hungarian ASR.

As for non-spontaneous (read) Hungarian databases such as (Roach et al. 1996; Pollák et al., 2000; Siemund at al, 2000), either size, composition or accessibility prevented their wider application for training and evaluation of ASR systems.

### 2.2 Spontaneous Hungarian ASR Results

Only a few attempts coping with the difficult task of recognizing spontaneous speech in this agglutinating language were published on some of the previously mentioned databases: MALACH (Mihajlik et al., 2010), HuComTech (Szaszák et al., 2011), and BEA (Beke and Szaszák, 2016). In short, the reported word error rates (WER's) are high, in the range of 42% – 55%, and it is difficult to draw valid conclusions about the effectiveness of the techniques applied throughout the studies.

---
[1] https://phon.nytud.hu/bea

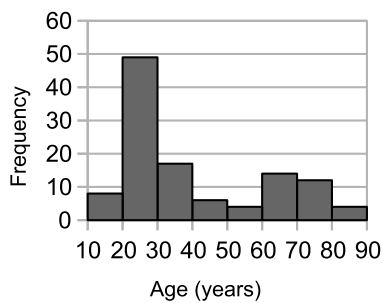 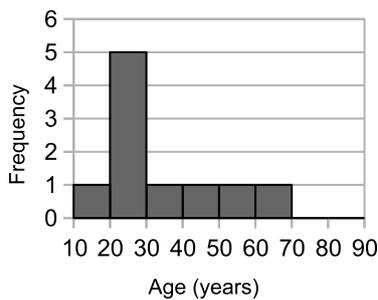 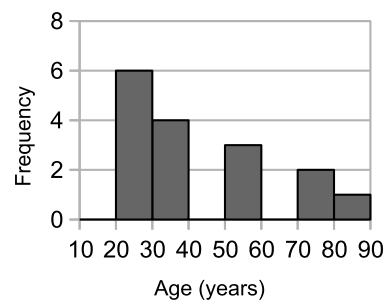

Figure 1: Age distribution of train-114.  Figure 2: Age distribution of dev.  Figure 3: Age distribution of eval.

|  | **train-114** | **dev-repet** | **dev-spont** | **eval-repet** | **eval-spont** |
|---|---|---|---|---|---|
| length [hours] | 71.2 | 0.65 | 4.02 | 0.95 | 4.91 |
| num of segments | 76 881 | 568 | 4 893 | 858 | 5 693 |
| num of characters | 3 104 165 | 28 467 | 154 994 | 43 448 | 197 738 |
| num of words | 555 322 | 4 110 | 27 939 | 6 229 | 35 178 |
| OOV words [%] | – | 2.0 | 7.3 | 1.4 | 7.9 |
| 3-gram ppl | – | 43.7 | 283 | 44.6 | 317.4 |

Table 1: Statistics of BEA-Base

## 3. The BEA-Base Database and Task

### 3.1 About the BEA Database

The original BEA ("BEszélt nyelvi Adatbázis" in Hungarian, meaning spoken language database) aimed at collecting speech data from 500 speakers, representative in age, sex, dialect, and educational background, primarily for linguistic research purposes. The recording sessions with 1 hour target length followed a protocol where the following modules were subsequently applied to each participant:
- Introduction of the speaker (job, family, hobby, etc.) – spontaneous monologue-like speech.
- Sentence repetitions – the interviewer reads, the subject repeats fixed, phonetically rich sentences.
- Opinion about a topic given by the interviewer – spontaneous monologue-like speech.
- Oral content summarization of two short stories told previously to the subject – again, monologic speech, spontaneous style.
- Free three-party conversation on a new topic.
- Reading the previously repeated fixed sentences.
- Reading a coherent text.

For more detailed description, see (Gósy, 2013; Neuberger et al., 2014). So far, speech data of 470 speakers have been recorded in studio environment. The acoustic conditions and digital processing (1 channel, sampling frequency of 44.1kHz, 16-bit linear quantization) remained constant during the whole process of data acquisition. The annotation/transcription procedure is evolved from typing into MS Word to using Transcriber[2] and then Praat[3], and currently preliminary ASR transcripts are being corrected manually using crowd sourcing tools. The transcription of the BEA is still in progress. For this reason and because of the heterogeneity of annotation formats and styles of the BEA database it is not yet suitable for machine learning purposes.

### 3.2 Construction of BEA-Base

To make use of the BEA database in ASR, we selected the most concise part of the dataset which was originally annotated using the Transcriber software tool. We kept only the 'core' of the transcription, the words in plain text, and all other notations (e.g., hesitations, noises) were deleted. Based on manual segmentation, all segments containing unintelligible or parallel speech were excluded from further processing. Speech data was resampled to 16kHz, 16 bits.

Speech data of 114 speakers was assigned to the training set, 10 for development 16 for evaluation sets. The ratio of Male/Female speakers in the train set is close to 40% whereas the dev and eval sets are perfectly balanced with a ratio of 50%. The age distributions are displayed in Figure 1–3.

The recordings, by default, would contain both the speech of the interviewer and the discourse partner. Since the identity of the experiment leader and her partner did not change in each recording session, we had to remove their speech segments from the dev end eval sets to ensure speaker independency in tests. According to the speech style, we separated dev and eval sets into spontaneous and non-spontaneous subsets. The latter one is identified as "repetitional" – or "repet" in short – because it contains repeated and read sentences by the subject based on the same written sentences. For meaningful test results, therefore, the full "repeated sentences" and "read sentences" blocks were excluded from the training set. For statistics, see Table 1.

### 3.3 BEA-Base ASR Task

Basically, our aim was to minimize the WER (Word Error Rate) on the spontaneous evaluation set – the results on the fixed, repetitive set are reported merely as references for non-spontaneous speech. By default, the train-114 set, i. e. the waveforms and transcriptions included were being used for the parameter estimation (training) of either acoustic, language or end-to-end models. However, because of the relatively small size of BEA-Base and because of the multitude of other language resources and public models available, we extended our investigations towards cross-language transfer learning, including semi-supervised pretraining. In all the experiments, the "dev-spont" set was used for validation (for optimizing hyperparameters). For the reproducibility of the results, we applied only open-source toolkits and publicly accessible models/language resources in transfer learning.

---
[2] http://trans.sourceforge.net/
[3] https://www.fon.hum.uva.nl/praat/

| Structure / num of param. | Unit | dev-repet | dev-spont | eval-repet | eval-spont |
|---|---|---|---|---|---|
| TDNN-F / 18M | phoneme | 7.06 / 1.82 | 27.12 / 9.04 | 6.26 / 1.58 | 28.41 / 9.36 |
|  | character | 7.30 / **1.70** | 27.08 / 8.90 | **6.08 / 1.51** | 28.28 / 9.26 |
| CNN-TDNN-F / 16M | phoneme | 7.10 / 1.81 | 27.01 / 8.95 | 6.33 / 1.62 | 28.81 / 9.52 |
|  | character | **6.98** / 1.77 | **26.71 / 8.59** | 6.28 / 1.59 | **28.15 / 9.13** |

Table 2: Kaldi based WER[%] / CER[%] results on BEA-Base

## 4. Hybrid HMM-DNN Baselines

One of the most successful ASR toolkits is undoubtedly Kaldi launched in 2011 (Povey et al., 2011). It offers a highly efficient WFST-HMM (Weighted Finite State Transducer – Hidden Markov-Model) framework and sophisticated DNN (Deep Neural Network) based acoustic modeling. In the experiments, we selected the most advanced "chain" approach with LF-MMI (Lattice Free Maximum Mutual Information) criteria and TDNN-F (Time Delay Neural Network-Factored) neural structure applying left-biphone shared-state models (Povey et al., 2018). Also the extension of the TDNN-F technique with Convolutional Neural Networks (CNN) (Zorila et al, 2019) was investigated.

Although Kaldi does support phonemic representations of words by weighted pronunciation alternatives, our earlier studies (Mihajlik et al., 2010) showed the viability of character (grapheme) based acoustic modeling for Hungarian, so we applied both approaches.

The standard s5 WSJ (Wall Street Journal) recipe was used throughout the experiments, with minor modifications. First, we added ±10% speed perturbations (only) as data augmentation. The other minor difference was that no i-vectors were applied. Word 3-grams were trained on the train-114 transcription data according to the (Chen and Goodman, 1999) method by using the SRILM toolkit. See Table 1. for text statistics. Beyond WER, Character Error Rates (CER) are also reported with respect to the agglutinative nature of Hungarian.

As Table 2 shows, it is much more challenging to transcribe automatically spontaneous speech than read/repeated speech. It is though not clear, whether acoustic or language model level difficulties can be attributed more to this phenomenon – elevated perplexities clearly indicate the existence of the latter one (see Table 1). Otherwise, we can observe only marginal differences between the approaches confirming the validity of character (or grapheme) based acoustic modeling for Hungarian. The best results are marked in bold.

## 5. End-to-end Neural Network Baselines

Proposed earlier by (Graves és Jaitly, 2014; Hannun et al., 2014), recently the most popular approach to tackle automatic speech to text conversion is to use deep neural network, from processing the input signal (such as Mel-Spectrogram) to the very end of text generation. There are many toolkits available today, e.g., ESPNet (Watanabe et al., 2018), Espresso (Wang et al., 2019), NVIDIA NeMo (Kuchaiev et al., 2019), and SpeechBrain (Ravanelli et al., 2021). In this study, we selected the last two for making baseline experiments on BEA-Base.

### 5.1 Convolutional Models

As (Zeghidour et al. 2018) showed, applying sequential 1D convolutions – with normalizations and residual connections – can be effective in ASR not only in terms of accuracy but in computational efficacy. (Kriman et al., 2020) introduced time-channel separated convolutions reducing the number of model parameters significantly.

#### 5.1.1 Training from Scratch

In the experiments we followed the recipe of (Kriman et al., 2020) applying the same QuartzNet (encoder) structure consisting of BxR consecutive convolutional blocks, a shallow decoder with CTC loss (Graves et al., 2006) and character output labels. Similarly, speed perturbation and SpecAugment (Park et al., 2019) was used for data augmentation and long, 1200 epochs training. By default, we used a batch size of 32 per GPU in a 2–4 A6000 graphic accelerators setup. All other hyperparameters of (Kriman et al., 2020) were kept, only the learning rate was optimized for each specific structure. Beyond greedy decoding, we applied a beam-search decoder for LM rescoring, with a beam size of 80. The same word 3-gram language model was used as in the previous experiments. The WER and CER results can be seen in Table 3 when **only BEA-Base** was used as a language resource for the estimation of QuartzNet model parameters. The best numerical results achieved so far are marked in bold.

In opposite to (Kriman et al., 2020), where the 5x3 structure was found optimal in the similar size WSJ task, it performed poorly in our experiments as compared to the 15x1 structure, having practically the same number of parameters. The best results were obtained with the 15x3 structure – we could not yet explore why the 15x5 version gave higher error rates; performing more extensive hyperparameter optimization might have achieved better performance.

| Structure / num of param. | LM | dev-repet | dev-spont | eval-repet | eval-spont |
|---|---|---|---|---|---|
| 5x3 / 6.4M | – | 23.82 / 5.38 | 36.16 / 13.24 | 27.85 / 6.56 | 38.30 / 13.64 |
| | 3-gram | 8.91 / 2.93 | 31.63 / 14.11 | 10.05 / 3.71 | 32.91 / 14.77 |
| 15x1 / 6.5M | – | 13.72 / 2.99 | 28.33 / 9.36 | 17.00 / 3.87 | 29.59 / 9.70 |
| | 3-gram | 6.76 / 2.01 | 26.66 / 10.45 | 6.90 / 2.43 | 27.97 / 11.29 |
| 15x2 / 9.6M | – | 10.05 / 2.42 | 26.02 / 8.58 | 12.49 / 3.12 | 27.22 / 9.06 |
| | 3-gram | 6.57 / 1.95 | 25.52 / 10.22 | 6.71 / 2.42 | 27.09 / 10.93 |
| 15x3 / 12.7M | – | 9.73 / 2.20 | **25.20** / **8.33** | 11.56 / 2.91 | **26.70** / **8.84** |
| | 3-gram | **6.50** / 1.86 | 25.50 / 10.0 | 6.86 / 2.36 | 26.83 / 10.76 |
| 15x5 / 18.9M | – | 12.70 / 2.89 | 26.43 / 8.42 | 13.79 / 3.32 | 27.63 / 8.96 |
| | 3-gram | 7.30 / 2.20 | 25.58 / 10.25 | 6.90 / 2.38 | 26.98 / 10.71 |

Table 3: QuartzNet based WER[%] / CER[%] results on BEA-Base

According to the expectations, adding a language model (LM) helps in the case of the less accurate end-to-end models and especially in the case of the repeated sentences where the perplexities are low. However, in case of the best model structure, using LM slightly increases the error values. This might be related to the large convolution kernel sizes: through them, the CTC-based end-to-end model perceives a large context with several word length output character labels. So, the end-to-end acoustic model does have some language modeling ability, and therefore an external language model trained on the same corpus cannot always improve the results, or, in our case, even deteriorates them by inhibiting the recognition of OOV words, for example.

### 5.1.2 Cross-Language Transfer Learning by Supervised Pretraining

In deep neural networks, starting training from scratch – as in the previous experiments – means randomly initialized weights (model parameters). Initializing weights with a pretrained model may give, however, better results even if the model is taught in another language. Compared to NLP, transfer learning across languages may be more effective in ASR, due to the similar use of speech organs in most of the languages.

Therefore, based on (Huang et al., 2020), we investigated the effect of borrowing the initial encoder weights from various language pretrained models. In the experiments English, German, and Italian language QuartzNet 15x5 models – available through NGC[4] – were used to initialize training on the Hungarian BEA-Base dataset. The learning rate was set to 0.002, the total batch size to 64 and the number of epochs to 300.

**Transfer learning results** with the QuartzNet 15x5 structure are shown in Table 4. (Note that no pretrained models are available with other structures, such as 15x3.) The model pretrained on English data clearly outperforms any previous or other language approaches on the spontaneous sets, and gives competitive accuracies on repeated speech, as well. German and Italian, although intuitively closer to the acoustics of Hungarian, both gave poorer results. A possible explanation may be that they were also initialized by the English pretrained model and then were extensively retrained (fine-tuned) on much smaller (CV 6.0) corpora. Consequently, they may represent the corpus used for fine tuning more than the data applied for pretraining, and so the model generality is reduced. Eventually, both languages provide better results as compared to initialization with random weights (see

| Pretraining Language | Data size [hours] | LM | dev-repet | dev-spont | eval-repet | eval-spont |
|---|---|---|---|---|---|---|
| English | 3k | – | 8.93 / 1.96 | **23.55 / 7.55** | 10.63 / 2.58 | **24.87 / 7.96** |
| | | 3-gram | **5.99 / 1.66** | 24.29 / 9.52 | **5.83** / 1.92 | 25.23 / 9.62 |
| English » Italian | 3k + 160 | – | 10.92 / 2.39 | 23.78 / 7.88 | 11.91 / 2.93 | 25.24 / 8.34 |
| | | 3-gram | 6.28 / 1.81 | 24.43 / 9.50 | 6.39 / 2.24 | 25.84 / 9.89 |
| English » German | 3k + 700 | – | 11.48 / 2.35 | 24.42 / 7.89 | 13.12 / 2.98 | 25.67 / 8.35 |
| | | 3-gram | 6.18 / 1.79 | 24.78 / 9.97 | 6.20 / 2.10 | 26.09 / 10.32 |

Table 4: QuartzNet based transfer learning WER[%] / CER[%] results on BEA-Base

---
[4] https://catalog.ngc.nvidia.com/models

| Num of param. | LM | dev-repet | dev-spont | eval-repet | eval-spont |
|---|---|---|---|---|---|
| 133M | – | 9.80 / 3.65 | **24.68** / 9.01 | 6.61 / 1.92 | **25.21** / 9.46 |

Table 5: CRDNN+GRU+CTC+Attention+BPE_600 based WER[%] / CER[%] results on BEA-Base

Table 3), confirming the effectivity of cross-language transfer learning.

## 5.2 Sequence-to-sequence Models with Attention

A theoretical limitation of CTC (and HMM) based ASR is the conditional independence assumption in the calculation of output symbol probabilities. To overcome this, (Chan et al., 2016) introduced an attention layer between the encoder and decoder modules of the neural network. (Watanabe et al., 2017) improved end-to-end models further by applying a joint CTC-Attention model based multi-task learning. In the next phase of the experiments, we used the SpeechBrain toolkit (Ravanelli et al., 2021) that supports sequence-to-sequence ASR models – including recurrent and transformer structures – and joint CTC-attention training.

### 5.2.1 Training from Scratch

First, we limited the language resources exclusively to the BEA-Base and applied the CRDNN (Convolutional, Recurrent and Deep Neural Nets) encoder and GRU (Chung et al., 2014) decoder modules connected by embedding and attention layers. We trained a BPE (Sennrich et al., 2015) tokenizer on the BEA-Base transcription with a vocabulary size of 600. The end-to-end neural model was trained for 60 epochs using the joint CTC-attention loss. All other hyperparameters were derived from SpeechBrain's CommonVoice recipe. No external language model was applied. During the evaluation, we used a beam size of 80 – similarly to the previous experiments. The **BEA-Base only** results can be seen in Table 5.

By comparing Table 3 and Table 5, we can observe a slight improvement regarding spontaneous data (marked in bold) and on the repetitive evaluation set. The error rates, however, are significantly higher than the ones obtained in the previous transfer learning experiments (see Table 4). Therefore, we continued our investigations in this direction.

### 5.2.2 Self-Supervised Pretraining based Transfer Learning

Since the introduction of transformer (Vaswani et al., 2017), it is used widely for sequence learning, including end-to-end ASR. One of the most popular ASR application of transformers is the wav2vec 2.0 encoder approach (Baevski et al., 2020) mainly due to its self-supervised contrastive learning ability inspired by the BERT model (Devlin et al., 2018). Therefore, in the subsequent experiments we replaced the encoder module to the wav2vec2-large transformer structure with 320 million parameters, by using SpeechBrain's corresponding CV recipe. This time, we did not train from scratch – the amount of data in BEA-Base was clearly insufficient for this. Instead, we applied several pretrained wav2vec2.0 (large) models available through HuggingFace[5] (see Table 7) for encoder weight initialization. Some of the pretrained models had been already fine-tuned in a target language, but in our perspective, this fine-tuning step is just a second (retraining) phase of pretraining by using supervised data. Unlike in the previous experiments in Section 5.1.2, the first (or only) phase of pretraining used here does not require any transcription text. Note that because the applied self-supervised pretraining relies entirely on the acoustic signal, there is no theoretical difficulty in doing it on multilingual corpora as has been done in (Conneau et al., 2021; Babu et al., 2021). We again trained the neural models for 60 epochs with a total batch size of 16 and left all other hyperparameters unchanged.

As Table 6 shows, self-supervised pretraining based acoustic transfer learning reduces the error rates dramatically, especially on the most challenging spontaneous sets. In this setup, Italian (Wang et al., 2021) and English (Baevski et al., 2020) based results are on the same level even if the second one was pretrained with a magnitude more unsupervised data. This time, German[6] based error rates are, in general, lower than in the case of the previous languages, however, this may be attributed more to the multilingual self-supervised pretraining than to the supervised retraining to German. Looking at the error rates of Turkish[6] and Hungarian[6] based approaches, we can observe similar accuracies to German[6], suggesting the priority of the self-supervised pretraining – which was identical in the case of these three languages.

| Pretraining | | dev-repet | dev-spont | eval-repet | eval-spont |
|---|---|---|---|---|---|
| Language | Data size [hours] | | | | |
| English | 60k | 8.61 / 3.19 | 18.01 / 5.45 | 8.46 / 2.59 | 19.17 / 5.94 |
| Italian | 4.5k | 8.18 / 3.32 | 17.74 / 5.61 | 7.45 / 2.50 | 19.07 / 6.44 |
| Multi » German | 53k + 700 | 6.42 / 1.31 | 17.93 / 6.89 | 6.66 / 2.25 | 17.99 / 5.55 |
| Multi » Turkish | 53k + 20 | 6.25 / 1.35 | 17.27 / 5.56 | 7.50 / 2.46 | 18.06 / 5.53 |
| Multi » Hungarian | 53k + 33 | 5.89 / 1.20 | 15.80 / 4.94 | **5.60** / 2.63 | 17.00 / 5.45 |
| | | | | | |
| Multi | 53k | **5.09 / 1.12** | 16.24 / 5.17 | 5.81 / 2.09 | 16.62 / 5.53 |
| Mega | 440k | 5.28 / 1.15 | **14.95 / 4.70** | 6.16 / 2.39 | **15.61 / 5.11** |

Table 6: wav2vec2-large+GRU+CTC+Attention+BPE_600 based transfer learning WER[%] / CER[%] results on BEA-Base

---

[5] https://huggingface.co/models

[6] For huggingface model ID and details, see Table 7.

| En | facebook/wav2vec2-large-lv60 |
|---|---|
| It | facebook/wav2vec2-large-it-voxpopuli |
| M»Ge | jonatasgrosman/wav2vec2-large-xlsr-53-german |
| M»Tr | m3hrdadfi/wav2vec2-large-xlsr-turkish |
| M»Hu | jonatasgrosman/wav2vec2-large-xlsr-53-hungarian |
| M | facebook/wav2vec2-large-xlsr-53 |
| Mega | facebook/wav2vec2-xls-r-300m |

Table 7: Huggingface ID's of pretrained models

Finally, it is easy to notice the superiority of results obtained by the direct fine-tuning of the large-scale (Conneau et al., 2021) "Multi" and very large-scale (Babu et al., 2021) "Mega" multilingually pretrained models. Apparently, the size – in terms of the amount of unlabeled audio training data – does matter, as the "Mega" set based approach, pretrained on 440 000 hours of speech, clearly gives the overall best results on the most relevant spontaneous sets. The xlsr-53k or "Multi" model, trained on smaller, 53 000 hours of multilingual data still performs well, competitive on the repetitional subsets. Interestingly, the retraining of this "Multi" model on Hungarian CV and audiobook data (see Table 6, "Multi » Hungarian" model) gave, in average, worse results than the direct fine-tuning of the original "Multi" model on BEA-Base. Our conclusion is that self-supervised pretraining is a must (if affordable) and one should start with a large-scale multilingual pretrained model for Hungarian ASR – and maybe for other languages, as well. Overall best results on Table 6 are marked in bold.

## 6. Conclusions

As an effect of the "deep learning revolution", language and speech technology made a truly remarkable progress in the past decade. Without accessible language resources and benchmarks, however, it is difficult to profit from the results. In automatic speech recognition, to date, one of the most important topics is the accurate transcription of spontaneous speech, and therefore a matching benchmark dataset is essential. In this paper we introduced the BEA-Base built purposely for this task. We showed that a heterogenous speech database designed originally for linguistic research can be turned – at least partially – into a concise dataset applicable for machine learning, primarily for ASR training and evaluation in Hungarian. By keeping the manual segmentations of natural spontaneous speech but removing the discourse partner's speech, a conversation with an AI assistant is mimicked. Therefore, the test sets are particularly suitable for evaluating ASR for conversational AI applications. The experiments show that accurate transcription of real spontaneous speech is much more challenging than in the case of read/repeated speech – even if there is no simultaneous speaking or significant background noise. The results demonstrate that despite the relatively small size of the training data (70 hours) it is possible to train ASR models solely on BEA-Base. Allowing other language or multilingual pretraining reduces the errors substantially. Using open-source tools and public pretrained models a WER of 15.61% and a CER of 5.11% is achieved on the spontaneous evaluation set, without using any supervised data or LM beyond the BEA-Base. In the development of baselines, we did not strive for the highest accuracies possible, but to provide solid baselines that can be reproduced easily. The database and the checkpoints of the best ASR baselines have been made available for the research community.

As for future work, regarding ASR evaluations, our plan is to continue investigations by adding external language model – the challenge here is to produce a matching spoken language text corpus with a sufficient size. In terms of ASR database development, we plan to release a BEA-Large version as soon as possible. Besides, a multi-level Praat TextGrid annotation is in progress which will provide an even more valuable language resource for linguists and technologists, as well.

## 7. Acknowledgements

Our work was supported by the following grants: NKFI-135038, NVIDIA Academic Hardware Grant 2021, ITM-NKFIH MILab.

## 8. Bibliographical References

## 9. Language Resource References